\documentclass[%
superscriptaddress,
preprint,
bibnotes,
 amsmath,amssymb,
 aps,
prb,
floatfix,
]{revtex4-1}

\usepackage{graphicx}
\usepackage{dcolumn}
\usepackage{bm}
\usepackage{float}
\usepackage{color}
\usepackage{mathbbol} 
\usepackage{subfigure}
\floatstyle{plain}
\usepackage{amsmath,amsxtra,bm,mathcomp,textcomp}

\providecommand{\Trace}[1]{\ensuremath{\text{Tr}\{\,#1\,\}}} 

\newcommand{\uncomment}[1]{}
\providecommand{\lr}[1]{\ensuremath{\langle #1 \rangle}}

\providecommand{\fa}[1]{f^{\ast}_{#1}}
\providecommand{\ca}[1]{c^{\ast}_{#1}}
\providecommand{\cao}{c^{\ast}}
\providecommand{\fao}{f^{\ast}}
\providecommand{\gm}[1]{g^{-1}_{#1}}

\definecolor{red}{rgb}{1,0,0}
\definecolor{green}{rgb}{0,1,0}

\begin{document}
\newfloat{example}{t}{lop}
\floatname{example}{Example}
\title{Dual-Fermion approach to Non-equilibrium strongly correlated problems.}

\author{C. Jung}
\affiliation{I. Institute of theoretical Physics, University of Hamburg, 20355 Hamburg, Germany}
\author{A. Lieder}
\affiliation{I. Institute of theoretical Physics, University of Hamburg, 20355 Hamburg, Germany}
\author{S. Brener}
\affiliation{Max Planck Research Department for Structural Dynamics, University of Hamburg-CFEL}
\author{H. Hafermann}
\affiliation{Max Planck Research Department for Structural Dynamics, University of Hamburg-CFEL}
\author{B. Baxevanis}
\affiliation{I. Institute of theoretical Physics, University of Hamburg, 20355 Hamburg, Germany}
\author{A. Chudnovskiy}
\affiliation{I. Institute of theoretical Physics, University of Hamburg, 20355 Hamburg, Germany}
\author{A. N. Rubtsov}
\affiliation{Department of Physics, Moscow State University, 119992 Moscow, Russia}
\author{M. I. Katsnelson}
\affiliation{Institute for Molecules and Materials, Radboud University of Nijmegen, 6525 AJ Nijmegen, The Netherlands }
\author{A. I. Lichtenstein}
\affiliation{I. Institute of theoretical Physics, University of Hamburg, 20355 Hamburg, Germany}

\date{\today}

\begin{abstract}
We present a generalization of the recently developed dual fermion approach introduced for correlated lattices to non-equilibrium problems. In its local limit, the approach has been used to devise an efficient impurity solver, the superperturbation solver for the Anderson impurity model (AIM). Here we show that the general dual perturbation theory can be formulated on the Keldysh contour.
Starting from a reference Hamiltonian system, in which the time-dependent solution is found by exact diagonalization, we make a dual perturbation expansion in order to account for the relaxation effects from the fermionic bath.
Simple test results for closed as well as open quantum systems in a fermionic bath are presented.
\end{abstract}

\maketitle

\section{Introduction}

In the last years there has been an evergrowing experimental as well as theoretical interest to non-equilibrium physics in strongly correlated systems.
Recent experiments \cite{wall2009ultrafast,tobey2008ultrafast} in the field of time resolved measurements have set tasks for theory that can not be fulfilled by the existing methods of analytical or numerical approaches to the considered systems.
Apart from the time-dependent DMRG\cite{White} and NRG\cite{Anders} methods, where a physical system is approximated by a finite one-dimensional chain, all other theoretical approaches are perturbative in a general sense, i.e. one subdivides the Hamiltonian of the system in two parts $H=H_0+H_{\mbox{pert}}$, solves the $H_0$ part exactly and takes the rest into account perturbatively.

Historically the first calculations of the transport through a quantum dot were the linear response investigation of the conductance by Landauer\cite{landauer1970electrical}. The further development of this kind of theory\cite{Schoen} lead to a series of strong coupling works. In those the perturbation expansion is done in powers of the tunneling amplitude (up to 2nd or 4th order) and the correlated part of the system is considered as an multi-orbital impurity. More sophisticated studies are based on the Keldysh technique. In the work of Wingreen and Meir \cite{wingreen1994anderson} a general formula for the time evolution of a system as a function of the dot Green's function and the tunneling is obtained and then the limit of weak symmetric tunneling is considered.

Another way to proceed is the weak-coupling approach, which starts from the transport through a non-interacting dot and applies Keldysh diagrammatics in terms of the Coulomb interaction. Analytical studies have been performed up to second order of the perturbation series and GW approximation\cite{thygesen2008impact}. Recent numerical studies based on a generalization of the continuous time Quantum Monte Carlo scheme (CT-QMC)\cite{Rubtsov2005weak} to the Keldysh contour \cite{werner2010weak} are, in principle, also a natural generalization of the weak-coupling perturbation theory, where diagrams are sampled stochastically. The CT-QMC approaches are, though in principle exact, limited to rather short time scales, since the calculations are strongly affected by a dynamical phase problem in the non-equilibrium case\cite{werner2010weak}.

Perturbative approaches all suffer from a common general drawback: they work well in some particular limit of parameter values, in which the system described by $H_0$ is a good starting point, and fail in the opposite limit. This is due to the fact that the unperturbed system in general is too simple to provide the basis of the perturbation theory in opposite limits.

One may instead consider the perturbation of a more complex system, simple enough to be solvable exactly and still sophisticated enough to include all important features of the considered full system so that all sources of complexity are possibly treated on equal footing. This task is in general non-trivial, as the systems that one would normally wish to take as the non-perturbed ones are strongly correlated and thus conventional diagrammatic techniques do not work.

The recently developed \cite{rubtsov2008dual} dual-fermion technique was originally intended to solve lattice problems beyond the well-established DMFT approximation, i.e. to incorporate a $k$-dependent into self-energies. One can consider it as an optimal perturbation around the DMFT starting point. In this technique, problems are solved in the above sense, i.e. a simple albeit nontrivial (reference) system
is solved exactly, and the dual perturbation theory accounts for the difference between the original and the reference system. Therefore this technique has been called a superperturbation. By "exactly" we mean finding the Green's function and, in principle, all higher cumulants.
In later works \cite{ldfa,hafermann2009superperturbation} it has been shown that the dual perturbation theory is indeed convergent in opposite limits and provides a sensible interpolation between these limits. For the case of the AIM, the dual perturbation theory has been to shown to pass into standard perturbation theory for the case of strong hybridization and weak interaction $U$, while it reproduces the hybridization expansion in the opposite limit of weak hybridization and strong interaction. We expect a corresponding behavior for the approach considered here.

In this work we generalize the dual fermion approach to models out of equilibrium and present an implementation for the experimentally important case of a hopping quench.
We use the path integral approach on the Keldysh contour to formulate the theory and to introduce a proper discretization scheme. We also present a simplified version of this method that is much less time- and computational power demanding but still proves to give qualitatively and sometimes even quantitatively correct results. The basic idea of the method is to generalize the previously discussed superperturbation impurity solver for the AIM. One replaces the continuous conduction electron bath by a small number of discrete bath levels (the reference system), solves this finite system by exact diagonalization (ED) and then accounts for the difference of hybridizations through the dual perturbation.
In the non-equilibrium technique, the idea remains the same, only that now we work on the Keldysh contour and all the relevant correlation functions depend on the time variables themselves and not on time differences as, of course, in the non-stationary situation there is no time translation invariance.

The remainder of this work is organized as follows: In section II we derive the discretized action of a general system on the Keldysh contour and perform the dual decomposition. In section III we give the details of the implementation of the algorithm and present some benchmark results, comparing them to other methods and discussing them. In the last section we summarize our work.

\section{Description of the method}
\subsection{Path integral formalism}
\begin{figure}[t]
\centering
\includegraphics[scale=0.7]{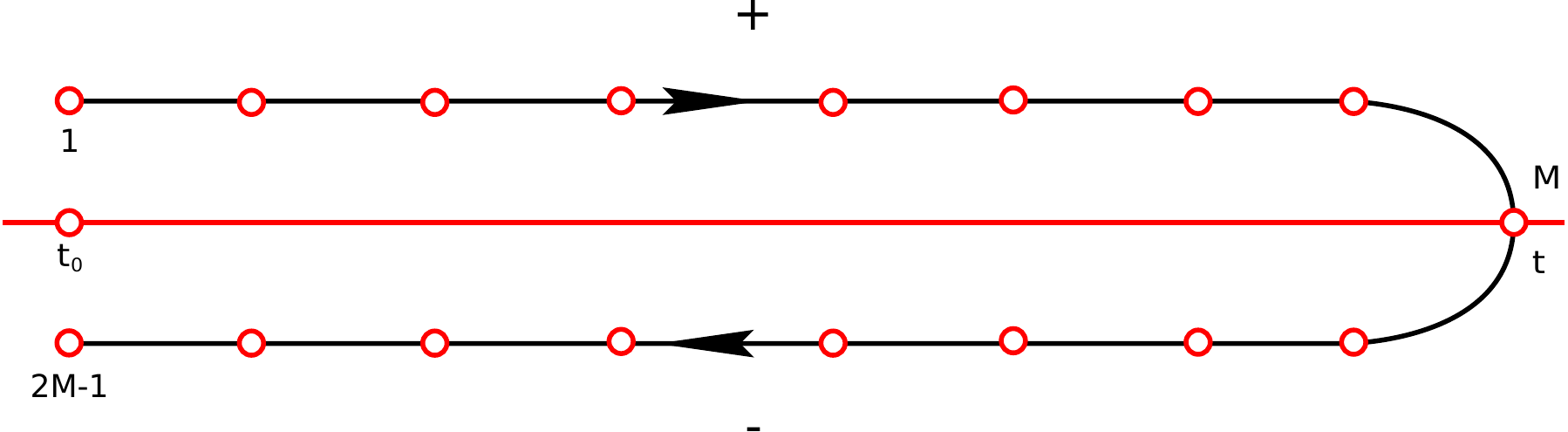}
\caption[Keldysh contour]{(Color online) Illustration of the Keldysh contour: the contour starts and end at $t_{0}$. Times are ordered in such a way, that points on the lower branch (-) are always later than points on the upper branch (+). }
\label{chap6:fig:keldcont}
\end{figure}

It is well-known \cite{Keldysh64,Keldysh65} that an accurate closed diagrammatic description of a non-equilibrium process is possible only on the so-called Keldysh contour Fig. \ref{chap6:fig:keldcont}. Here one propagates the state of the system first in the direction of increasing time along the time axis and then exactly backwards to the initial state. The reason for this is that in order to calculate averages of operators one has to propagate back to the initial state since the final state of the system, unlike in the zero-temperature ground state technique, is a priori unknown. Our goal will thus be to introduce the dual transformation for the Keldysh action.

In equilibrium quantum mechanics every observable $O$ is associated with a Hermitian operator $\hat{O}$. Its expectation value is given by $\langle \hat{O} \rangle=\Trace{\hat{O}\rho_{0}}$, where $\rho_{0}$ is the density matrix of the system governed by the Hamiltonian $H_{0}$. As long $\rho_{0}$ commutes with the Hamiltonian $[\rho_{0},H_{0}]=0$ the expectation value of $\hat{O}$ will not have any time dependence.

In the following we consider the situation in which the system is in equilibrium for times smaller than $t_0$ and is then perturbed by a sudden switch of some not specified internal parameter at $t=0$. The theory at hand is in principle able to treat general time dependent perturbations, but this requires a significant numerical effort.
For this case the expectation value of $\hat{O}$ is given by the average of $\hat{O}$ in the Heisenberg picture traced over the initial density matrix $\rho_{0}$:
\begin{equation}\label{chap6:init}
O(t)=\langle \hat{O}_{H}(t)\rangle=\Trace{\hat{O}_{H}(t)\rho_0}=\Trace{\hat{U}(t_0,t)\hat{O}\hat{U}(t,t_{0})\rho_{0}}
\end{equation}
$\hat{U}(t,t')$ is the evolution operator of the system. If $\hat{H}$ is the full, time dependent Hamiltonian of the system, the evolution operator is given by:
\begin{equation}\label{chap6:U}
\hat{U}(t,t')=
\begin{cases}
\hat{T}\exp{(-i\int_{t'}^{t}} d\bar{t} \hat{H}(\bar{t})) & t>t'\\
\hat{\bar{T}}\exp{(-i\int_{t'}^{t}}d\bar{t} \hat{H}(\bar{t})) & t<t'.
\end{cases}
\end{equation}
In the latter expression $\hat{T}$ is the time ordering operator which reshuffles the operators into chronological order, with earlier times to the right. The anti-chronological time ordering operator $\hat{\bar{T}}$ rearranges later times to the right.
Reading the time arguments of Eq. \eqref{chap6:init} from left to right, one sees that the time evolution of the observable starts at $t_{0}$, propagates to t, and goes back again to $t_{0}$ and hence can be depicted by the Keldysh contour shown in Fig. \ref{chap6:fig:keldcont}. Introducing a time-ordering operator  $\hat{T}_{C}$ which arranges operators according to their position on the contour, the time-evolution operator can be written in the following form:
\begin{equation}\label{chap6:contour_evol}
\hat{U}(t,t')=\hat{T}_{C}\exp{\left(-i\int_{t'}^{t} d\bar{t} \hat{H}(\bar{t})\right)},
\end{equation}
where the integral is taken along the contour. Considering the last part of Eq. \eqref{chap6:init}, the time evolution of the system can be viewed as an initial value problem. At time $t_0$ the system is prepared by an initial density matrix $\rho_0$ and then the time evolution of the system is governed by Eq. \eqref{chap6:contour_evol}. In cases where the system is correlated before time $t_0$ and the initial density matrix is unknown (i.e. $\rho_0$ cannot be written in the form $\exp{(B_0)}$, with some one-particle operator $B_0$), it is necessary to extend the contour in Fig. \ref{chap6:fig:keldcont} to imaginary times as described in reference\cite{WAGNER:1991fk}. This can be done straightforwardly, but for simplicity we restrict ourselves to the case of non-correlated $\rho_0$.

We start by deriving the Keldysh action following Kamenev\cite{2009AdPhy..58..197K} as it is most appropriate for our purposes. First we omit the interaction part and consider free electrons governed by the Hamiltonian $H(c^\dagger_\alpha,c_\beta)=\sum_{\alpha\beta}H_{\alpha\beta}c^\dagger_\alpha c_\beta$, where small Greek letters enumerate the single-particle states.

The Keldysh partition function can formally be defined as
\begin{equation}
Z=\Trace{\rho_{0}\hat{U}_{C}},
\end{equation}
which is a complex way to express $\Trace{\rho_{0}}$ through the evolution operator along the Keldysh contour. The next step is to discretize the contour with $2M-1$ points and insert the identity of the overcomplete fermion coherent-state basis \cite{Negele:1998fy}
\begin{equation}
 \Eins=\int\prod_{\alpha} d(c^{\ast}_{i\alpha},c_{i\alpha})\,e^{-\sum_{\alpha}c^{\ast}_{i\alpha}c_{i\alpha}}|c_i\rangle\langle c_i|
\end{equation}
at each discretization point $t_i$. Here $|c_i\rangle$ stands for coherent states corresponding to eigenvalues $c_{i\alpha}$, small Latin letters enumerate the discretization points. We obtain
\begin{align}
\begin{split}
Z=&\lim_{M\to\infty}\int\prod_{i=1}^{2M-1}\prod_{\alpha} d(c^{\ast}_{i\alpha},c_{i\alpha})e^{-\sum_{\alpha}c^{\ast}_{i\alpha}c_{i\alpha}}\langle-c_{1}|\rho_{0}|c_{2M-1}\rangle
\langle c_{2M-1}|\hat{U}(t_{2M-1},t_{2M-2})|c_{2M-2}\rangle \dots\\
&\qquad\qquad\times
\langle c_{M+1}|\hat{U}(t_{M+1},t_M)|c_{M}\rangle
\langle c_{M}|\hat{U}(t_M,t_{M-1})|c_{M-1}\rangle\dots\langle c_{2}|\hat{U}(t_2,t_1)|c_{1}\rangle
\label{action1}
\end{split}\\
=&\lim_{M\to\infty}\int\prod_{i=1}^{2M-1}\prod_{\alpha}d(c^{\ast}_{i\alpha},c_{i\alpha})e^{-\sum_{\alpha}c^{\ast}_{i\alpha} c_{i\alpha}} e^{-\sum_{\alpha\beta}c^{\ast}_{1\alpha}\rho_{0\alpha\beta}c_{2M-1\beta}}\nonumber\\
&\qquad\qquad\times \dots\times  e^{\sum_{\alpha\beta}c^{\ast}_{M+1\alpha}c_{M\beta}+i\Delta t H(c^{\ast}_{M+1\alpha}c_{M\alpha})}\nonumber\\
&\qquad\qquad\times e^{\sum_{\alpha\beta}c^{\ast\alpha}_{M}c_{M-1\alpha}-i\Delta t H(c^{\ast\alpha}_{M}c_{M-1\beta})}\times\dots \times e^{c^{\ast}_{2\alpha}c_{1\alpha}-i\Delta t H(c^{\ast}_{2\alpha}c_{1\beta})}
\label{action2}
\nonumber\\
=&\lim_{M\to\infty}\int\prod_{i=1}^{2M-1}\prod_{\alpha} d(c^{\ast}_{i\alpha},c_{i\alpha})e^{i \sum_{jj'\alpha\beta}c^{\ast}_{j\alpha}G^{-1}_{jj'\alpha\beta}c_{j'\beta}}
\end{align}
with $G^{-1}$ given by:
\begin{equation}\label{chap6:dis_action}
iG^{-1}_{ii'\alpha\beta}=\Delta t\left(-\delta_{\alpha\beta}\frac{\delta_{ii'}-\delta_{i,i'+1}}{\Delta t}\pm iH_{\alpha\beta}\delta_{i,i'+1}\right)-\delta_{i1}\delta_{i',2M-1}\rho_{0\alpha\beta}.
\end{equation}
In the latter expression the upper $(+)$ sign applies for $i>M$ and $(-)$ otherwise.

The transition from (\ref{action1}) to (\ref{action2}) in particular the transformation of the matrix element $(1,2M-1)$ uses the property of the coherent states:
\begin{equation}
\langle c|e^{c^\dagger_\alpha A_{\alpha\beta}c_\beta}|c'\rangle=\exp{(c^\ast_\alpha e^{t_{\alpha\beta}}c'_\beta)},
\end{equation}
which implies that $\rho_0$ is non-correlated. $\rho_{0\alpha\beta}$ in (\ref{action2}) is thus given by $\exp{(A_{\alpha\beta})}$ if the initial density matrix can be expressed as $\rho_0=\exp{(c^\dagger_\alpha A_{\alpha\beta}c_{\beta})}$.

The expression for $G^{-1}$ consists of three major parts: the first one describes the discrete time evolution backward in time along the lower branch of the contour, the second one the evolution forward and the third one represents the boundary condition of the fermionic states.

For reasons that will become apparent below, we will not take the limit $\Delta t\to0$, but seek for a suitable mapping of the Keldysh formalism on a time grid. Thus the above matrix form of the action is appropriate for us. The indices of the latter correspond to discretization points on the contour. It is not necessary to explicitly distinguish between the upper and lower branch of the contour and introduce the conventional Keldysh $+-$ indices. The inverse Green's function for two coupled non-interacting sites is shown in example \ref{chap6:example:two_site}.

Including of the interaction is straightforward, it results in terms of the type $Uc^*c^*cc$ in the action (which is valid in the limit $\Delta t\to0$). To finish of the preparation we notice that as baths are taken to be uncorrelated one can perform the Gaussian integration over the bath fields to obtain the action depending only on the impurity fields. This leads to the well-known hybridization function which in our case depends on two discrete time indices. Explicit calculation results in:
\begin{equation}\label{chap6:hyb}
 \Delta(t,t')=\sum_{t_1,t_2}\hat{V}(t,t_1)\hat{G}_{\text{bath}}(t_1,t_2)\hat{V}^{\dagger}(t_2,t').
\end{equation}
Here $\hat V$ and $\hat{G}_{\text{bath}}$ must be considered as supermatrices in time indices and bath degrees of freedom. The structure of a single block of $\hat V$ can be seen from example \ref{chap6:example:two_site}.
While in most works the grid is chosen in such a way that all points are equidistant and the number of points on the upper and lower contour is equal (see, e.g., Ref. \cite{freericks}), we here need to choose the dimension of the time block to be odd due to the structure of the dual perturbation theory (see below).
\begin{example}
\begin{align*}
iG^{-1}_{jj'}=
\begin{pmatrix}G^{-1}_{\text{imp}}&\hat{V}\\
\hat{V}&G^{-1}_{\text{bath}}
\end{pmatrix}=&
\left(
\begin{array}{ccc|ccc}
-1&0&\textcolor{red}{-\rho_{0}^{i}}&0&0&\textcolor{red}{-\rho_{0}^{ib}}\\
h_{+}&-1&0&v_{+}&0&0\\
0&h_{-}&-1&0&v_{-}&0\\
\hline
0&0&\textcolor{red}{-\rho_{0}^{bi}}&-1&0&\textcolor{red}{-\rho_{0}^{b}}\\
v_{+}&0&0&h_{+}&-1&0\\
0&v_{-}&0&0&h_{-}&-1\\
\end{array}
\right)
\end{align*}
\caption[Discrete version of an AIM action]{Inverse Green function for an interaction-free two site model. The Hamiltonian is given by $H=\epsilon_i c^\dagger c+\epsilon_b b^\dagger b+(Vc^\dagger b+\mbox{h.c.})$. Terms in the right upper edges (red) are a direct consequence of the anti-periodic bounding conditions on the time contour. The following abbreviations have been used: $v_{\pm}=\mp i V\Delta t$, $h_{\pm}=1\mp iH\Delta t$.}
\label{chap6:example:two_site}
\end{example}

\subsection{Dual perturbation theory on the Keldysh contour}
In this section we generalize the superperturbation approach for the AIM\cite{hafermann2009superperturbation} to the case of non-equilibrium systems. We start from the path integral formulation of the partition function:
\begin{equation}
Z=\int D[\ca{},c]\exp(iS[\ca{},c]),
\end{equation}
with the following discrete-time expression for the impurity action $S$ :
\begin{equation}
S[\cao,c]=\sum_{tt'}\sum_{ab}\ca{at}[G_{0tt'}^{-1}-\Delta_{tt'}]_{ab}c_{bt'}+S^{\text{U}}[\cao,c].
\end{equation}
Here $G_{0}$ is the Green function of the isolated impurity and $\Delta_{tt'}$ is the hybridization function, which describes the coupling of the impurity to a continuum of bath states.
To stress that we are working on a time grid, we write the indices of the time-dependent matrices explicitly. Indices for orbital and spin degrees of freedom have been summarized in Latin letters. $S^{\text{U}}[\cao,c]$ is some local interaction.\\

We introduce a reference system with the same interaction part as the original system, albeit with coupling to a set of discrete bath levels described by a hybridization function $\tilde{\Delta}_{tt'}$. For a small number of bath states, this system can be solved exactly using ED.
\begin{align}
\tilde{S}[\cao,c]=&\sum_{tt'}\sum_{ab}\ca{at}[G_{0tt'}^{-1}-\tilde{\Delta}_{tt'}]_{ab}c_{bt'}+S^{\text{U}}[\cao,c].
\end{align}
The action of the original system can be expressed in terms of the reference system and a correction. To this end, the hybridization of the reference system is added and subtracted:
\begin{align}
S[\cao,c]=&\tilde{S}[\cao,c] + \sum_{tt'}\sum_{ab}\ca{at}[\tilde{\Delta}_{tt'}-\Delta_{tt'}]_{ab}c_{bt'}.
\end{align}
One has to be aware that in the non-equilibrium case the action representation always involves boundary conditions that enter the discrete matrices as shown in Example \ref{chap6:example:two_site}. When adding and subtracting $\tilde{\Delta}_{tt'}$ we assume that both systems have the same $\rho$ contribution on the impurity. The other parts of the density matrix are less important, because the difference in these terms will be treated perturbatively by doing an expansion in $\tilde{\Delta}_{tt'}-\Delta_{tt'}$. In the equilibrium case it was not necessary to discuss the boundary conditions of the action, because they were automatically fulfilled when working with Matsubara frequencies.

Now dual fermions (represented by $\fa{}$ and $f$) are introduced through a Gaussian identity for Grassmann variables. The transformation can be written in the following form:
\begin{equation}
e^{\ca{1}n_{12}D^{-1}_{23}n_{34}c_{4}}=\frac{1}{\det{ D}}\int \mathcal{D}[\fa{},f]e^{-\fa{1}D_{12}f_{2}+\fa{1}n_{12}c_{2}+\ca{1}n_{12}f_{2}}
\end{equation}
with the following definitions for $n$ and $D$:
\begin{equation} \left.\begin{aligned}
n&=ig^{-1}_{12}\\
D&=ig^{-1}_{12}[\tilde{\Delta}-\Delta]^{-1}_{23}g^{-1}_{34}\end{aligned} \right\} \rightarrow n_{12}D^{-1}_{23}n_{34}=i[\tilde{\Delta}-\Delta]_{14},
\end{equation}
$g$ being the Green function of the reference problem.
After some straightforward algebra the partition function can be brought into the following form
\begin{align}
Z&=\exp \biggl(i\tilde{S}[\ca{},c]+i\sum_{tt'}\sum_{ab} \ca{at}[\tilde{\Delta}-\Delta]c_{bt'}\biggr)\nonumber\\
&=Z_{f}\exp\biggl(i\bigl\{ \fa{1}[-\gm{}(\tilde{\Delta}-\Delta)^{-1}\gm{}]_{12}f_{2}+\fa{1}\gm{12}c_{2}+\ca{1}\gm{12}f_{2}+\tilde{S}[\ca{},c]\bigr\}\biggr)\nonumber\\
&=Z_{f}\exp(iS[\ca{},c;\fa{},f]),
\end{align}
with
\begin{equation}
Z_{f}=\det (-ig[\tilde{\Delta}-\Delta]g).
\end{equation}
The action can be split into three parts: Two parts, which contain either $c$-fermions or dual variables, and a part that describes the coupling between them:
\begin{align}
S[\cao,c,\fao , f]&=\tilde{S}[\cao,c]+S^{\text{c}}[\cao , c ; \fao ,f]-\fa{1}[g^{-1}(\tilde{\Delta}-\Delta)^{-1}g^{-1}]_{12}f_{2},\\
\text{with:}&\nonumber\\
S^{\text{c}}[\cao , c ; \fao ,f]&=\fa{1}g^{-1}_{12}c_{2}+\ca{1}g^{-1}_{12}f_{2}.
\end{align}
In the last equation we have combined temporal, orbital and spin indices into numbers. To integrate out the $c$-fermion part, the following defining equation for the dual interaction potential $\mathcal{V}$ is introduced:
\begin{equation}\label{chap6:derivation:define_dual_pot}
\begin{split}
\int \mathcal{D}[\ca{},c] \exp\Bigl(i(\tilde{S}[\ca{},c]+S^{\text{c}}&[\ca{},c,\fa{},f] )\Bigr)\\
 &\overset{!}{=}\tilde{\mathcal{Z}}\exp\Bigl( i(-\sum_{12}\fa{1}g^{-1}_{12}f_{2} + \mathcal{V}[\fao,f])\Bigr).
\end{split}
\end{equation}
Expanding the left hand side in $S^{\text{c}}[\ca{},c;\fa{},f]$ and integrated over the $c$-fermion fields produces the correlation functions of the reference system.
The dual potential is evaluated by expanding the right hand side and comparing the expressions on both sides by order. Up to fourth order in $f$-fermion fields we obtain:
\begin{equation}
\mathcal{V}[\fao,f]=\frac14\gamma^{(4)}_{1234}\fa{1}f_{2}\fa{3}f_4+\dots\,.
\end{equation}
where $\gamma^{4}$ is the complete two-particle vertex of the reference problem:
\begin{equation}
\gamma^{(4)}_{1234}=g^{-1}_{11'}g^{-1}_{33'}(\chi_{1'2'3'4'}-\chi^{0}_{1'2'3'4'})g^{-1}_{2'2}g^{-1}_{4'4}\\
\end{equation}
with
\begin{equation}
\chi_{1234}=\langle c_1c^*_2c_3c^*_4\rangle
\end{equation}
being the full two-particle Green functions of the reference problem and
\begin{equation}
\chi^{0}_{1234}=g_{14}g_{32}-g_{12}g_{34}
\end{equation}
its disconnected part.
The dual action can now be written as follows:
\begin{equation}
S^{\text{d}}[\fa{},f]=\fa{1}(G^{\text{d}}_{0})^{-1}_{12}f_{2} +\frac{1}{4}\gamma^{(4)}_{1234}\fa{1}f_{2}\fa{3}f_{4}+\dots
\end{equation}
with the following definition of the bare dual Green's function:
\begin{equation}
G^{\text{d}}_{0}=-g[g+(\tilde{\Delta}-\Delta)^{-1}]^{-1}g.
\end{equation}
Now a usual diagrammatic expansion in powers of $\gamma$ can be performed. The lowest order contribution to the dual self-energy is thus
\begin{equation}
\Sigma^{\text{d}}_{12}\approx -i\gamma^{(4)}_{1234}G^{0\text{d}}_{43}.
\end{equation}

\begin{figure}[t]
\centering
\includegraphics[scale=0.6]{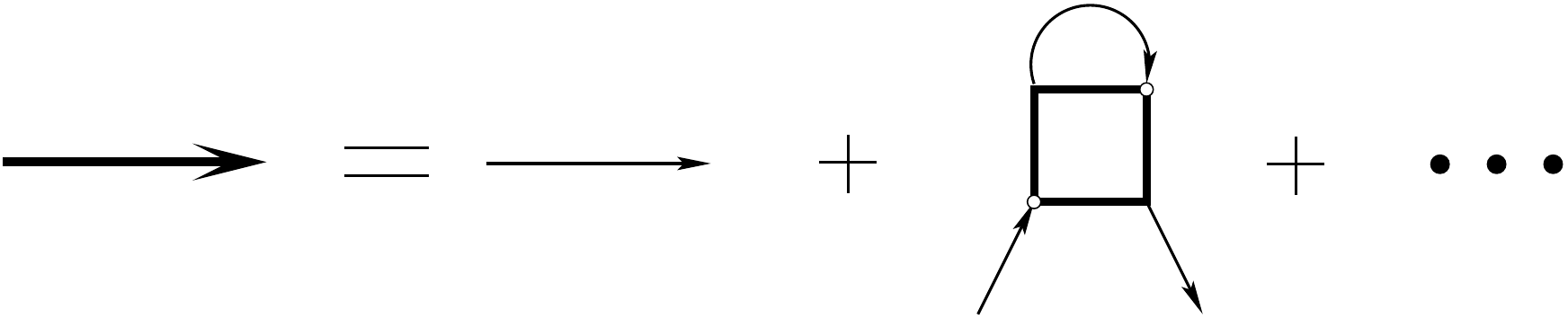}
\caption[Illustration of the lowest order contributions to the (dual) Keldysh Green's function]{Illustration of the lowest order contributions to the (dual) Keldysh Green's function: The full dual propagator is expressed in terms of the vertex and the bare dual propagator as lines.}
\label{chap6:green_def}
\end{figure}

The lowest order approximation to the Keldysh Green function is depicted diagrammatically in Fig. \ref{chap6:green_def}. After calculating an approximation to the dual propagator the result can be transformed back to c-fermions using the following exact relation:
\begin{equation}
 G=(\tilde{\Delta}-\Delta)^{-1}+[g(\tilde{\Delta}-\Delta)]^{-1}G^{d}[(\tilde{\Delta}-\Delta)g]^{-1}
\end{equation}

Note that the $g^{-1}$ factors in the definition of $\gamma^{(4)}$ at each vertex of a diagram cancel with the corresponding factors $g$ in the expression for $G^{\text{d}}_{0}$.

The above formulae require the inversion of the impurity Green function and hybridization. Therefore we employ a time discretization (note that, of course, the matrices are not diagonalized by Fourier transform).

Furthermore, there are some important side remarks for the numerical calculations. The first one involves the structure of the time grid. If it is chosen such that equally many equidistant points lie on the upper and lower branches of the contour and the total number of points is even, the time step between the last point on the upper contour and the next one on the lower contour is zero. This causes a vanishing matrix element in $\hat{V}$ and leads to a singular matrix for $\Delta(t,t')$. This can lead to a break down of the dual theory, because the hybridization $\Delta(t,t')$ has to be inverted. This problem can be cured if an additional point at the tip of the contour is introduced, which neither belongs to the upper nor to the lower contour. This additional point has no other consequences and can be treated without any further problems. That is the reason why unlike in the work \cite{2009AdPhy..58..197K}, the dimension of the time block in Example \ref{chap6:example:two_site} is odd.
Another issue in the inversion procedure can occur, if the off-diagonal elements $\rho^{ib}_{0}$ and $\rho^{bi}_{0}$ of the density matrix vanish. This happens, if the impurity is totally decoupled from the bath for times smaller than $t_0$. In that case $\hat{V}$ in Eq. \eqref{chap6:hyb} becomes also ill conditioned. This problem can be cured by introducing an infinitesimally small hopping parameter for times smaller  than $t_0$.

\subsection{Calculation of one- and two-particle Green's functions in exact diagonalization}
In the last section it has become clear that the key ingredients to construct the superperturbation theory are the one and two particle Green's function of the reference system. In the following we will give the formulae of those quantities in the Lehman representation.
We start with the single-particle Green's function. This quantity depends on two times and two indices:
\begin{equation}
 g_{\alpha\beta}(t,t')=-i\lr{T_{C}c_{\alpha}(t)c_{\beta}^{\dagger}(t')\rho_0}.
\end{equation}
To derive the spectral representation, identity matrices are inserted between the operators, and the time-evolution operator is written in its diagonal form:
\begin{equation}\label{chap6:ED:green}
\begin{split}
 g_{\alpha\beta}(t,t')=\frac{1}{Z_0}\sum_{0,i,j,k}&\left\lbrace-i\lr{0|i}\lr{i|c_{\alpha}|j}\lr{j|c^{\dagger}_{\beta}|k}\lr{k|0}e^{-\beta E_{0}}e^{i[E_{i}t+E_{j}(t'-t)-E_{k}t']}\cdot \theta_{C}(t-t')\right.\\
 &\left.+i\lr{0|i}\lr{i|c^{\dagger}_{\beta}|j}\lr{j|c_{\alpha}|k}\lr{k|0}e^{-\beta E_{0}}e^{i[E_{i}t'+E_{j}(t-t')-E_{k}t]}\cdot \theta_{C}(t'-t)\right\rbrace.
\end{split}
\end{equation}
$t$ and $t'$ are times on the Keldysh contour, which means that the $\theta$-functions are defined as follows:
\begin{equation}
 \theta_{C}(t'-t)=\begin{cases}
	      0 & \text{if $t>t'$}\\
              1 & \text{if $t<t'$},
              \end{cases}
\end{equation}
where the relation symbols order the times along the contour.\\
In the above, $i,j,k$ denote the exact eigenstates of the reference problem, whereas the states $|0\rangle$ account for the initial state of the system described by the density matrix $\rho_0$. As mentioned above, we consider only such cases where the initial density matrix can be given as $e^{-\beta B_0}$, where $\beta$ is some effective temperature and $B_0$ is a one-particle operator. Thus, states $|0\rangle$ are the eigenstates of this auxiliary operator and $E_0$ are the corresponding eigenvalues. $Z_0=\sum_0 e^{-\beta E_0}$

The formal definition of the two particle Green's function is given by the following expression:
\begin{align}
 \chi_{\alpha\beta\gamma\delta}(t_{1},t_{2},t_{3},t_{4})=& \lr{T_{C}c_{\alpha}(t_{1})c_{\beta}^{\dagger}(t_{2})c_{\gamma}(t_{3})c_{\delta}^{\dagger}(t_{4})\rho_0}\\
 =&\lr{T_{C}O_{1}O_{2}O_{3}O_{4}\rho_{0}}.
\end{align}
Here it was necessary to introduce abbreviations for the operators, in order to rewrite the time-ordered product as a sum over all possible permutations multiplied by a  $\theta$-function depending on the four time arguments:
\begin{equation}\label{chap6:two_particle_GF}
 \begin{split}
 \chi_{\alpha\beta\gamma\delta}(t_{1},t_{2},t_{3},t_{4})=& \frac{1}{Z_0}\sum_{\pi\in S_{n}}(-1)^{\pi}\theta_{C}(t_{\pi_1}>t_{\pi_2}>t_{\pi_3}>t_{\pi_4})\\
& \times\sum_{0,i,j,k,l,m}\lr{0|i}\lr{i|O_{\pi_1}|j}\lr{j|O_{\pi_2}|k}\lr{k|O_{\pi_3}|l}\lr{l|O_{\pi_4}|m}\lr{m|0}\\
&\times e^{i[E_{i}(t_{\pi_1})+E_{j}(t_{\pi_2}-t_{\pi_1})+E_{k}(t_{\pi_3}-t_{\pi_2})+E_{l}(t_{\pi_4}-t_{\pi_3})+E_{m}(-t_{\pi_4})]}\\
&\times e^{-\beta E_0}.
 \end{split}
\end{equation}
$S_n$ being the permutation group of the four time points.
Since the two-particle Green's function depends on four discrete time arguments, it is not feasible to store $\chi$ in computer memory. We therefore evaluate Eq. \eqref{chap6:two_particle_GF} on the fly every time it is needed in the perturbation expansion. This can be done most efficiently if one precalculates the time-dependent creation and annihilation operators and only performs the matrix product and the subsequent trace at each time combination. Since this procedure is based on standard matrix multiplication, the computational effort of a single evaluation of Eq. \eqref{chap6:two_particle_GF} scales as $O(n^{3}_{\text{max block}})$, where $n_{\text{max block}}$ is the dimension of the largest symmetry block in the Hamiltonian.
An alternative way to calculate the ED quantities is to calculate the time dependence of the eigenstates from the Schr\"odinger equation and storing the operator matrices in memory as a function of time \cite{eckstein}. The diagrams can then be evaluated using higher-order quadrature rules.
Depending on the number of time points it is possible to efficiently treat reference systems up to a total system size of at least four sites in total.

\section{Application of the method}
\subsection{Simple test}
\begin{figure}[t]
\centering
\subfigure[]{
\includegraphics[scale=0.9]{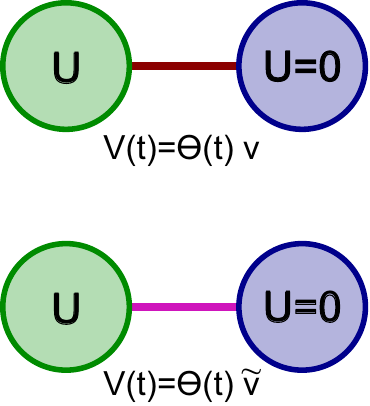}
\label{chap6:fig:model}
}\hspace*{0.8cm}
\subfigure[]{
\includegraphics[scale=0.8]{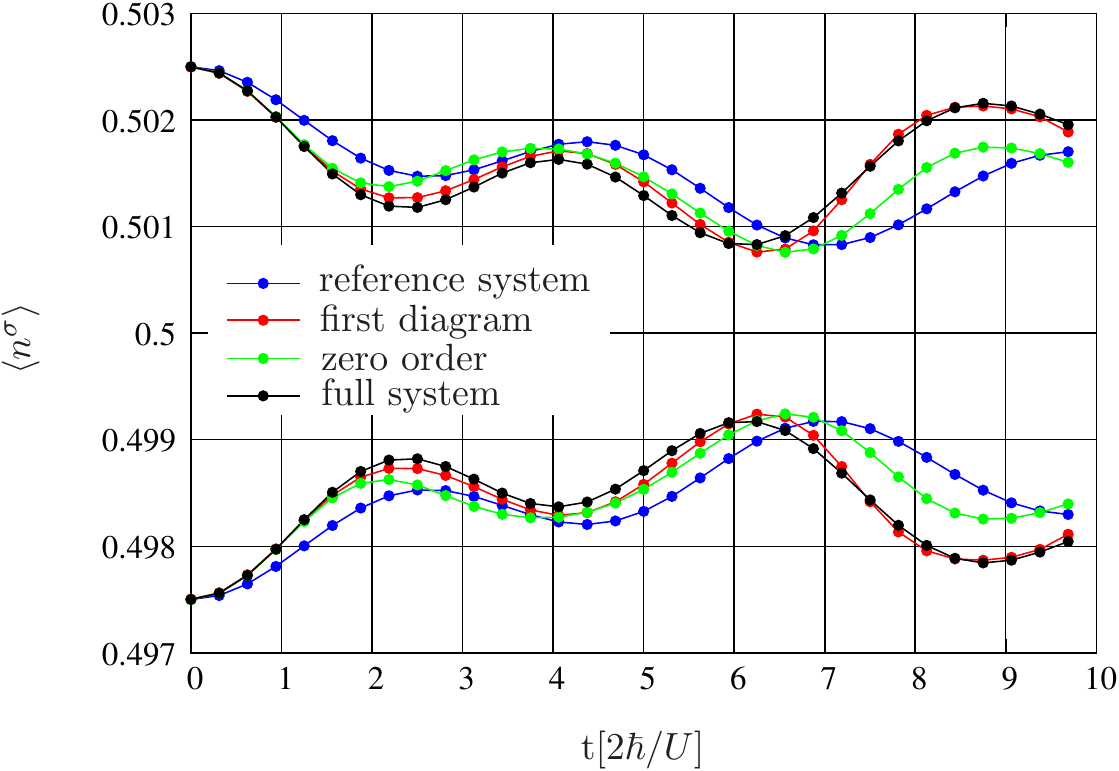}
\label{chap6:fig:short_time_plot}
}
\caption[First test of SPERT on the Keldysh contour ]{(Color online) Simple test of the superperturbation method on the Keldysh contour. (a) the model under consideration consists of one interacting site coupled to one bath site (upper left picture). The time dependence consists of a sudden switch in the hopping amplitude from an infinitesimally small value to a non-zero one. The reference system is prepared in the same way, but the hopping is switched to a lower value (lower left picture). (b) Plot of $n^{\sigma}(t)$ for the full system, the reference system and different degrees of approximation. The zero-order curve corresponds to a dual theory without any diagrams, the first-order curve to a solution including the first diagram. Both curves are shifted from the reference solution towards the exact result. The data points corresponding to the solution with the first diagram are in good agreement with the solution of the full system. The calculations have been done for the following parameters: $\beta=5$, $U=2$, $v=0.5$, $\tilde{v}=0.4$, $B=0.001$.}
\label{chap6:fig:first_test_keld_n_t.pdf}
\end{figure}

As a first test we calculate the time evolution of an exactly solvable model. Figure \ref{chap6:fig:model} shows the model under consideration.
The full system consists of an interacting site coupled to one additional bath site. At time $t_0$ the system is prepared in such a way that both sites are half filled and the spin degeneracy on the interacting site is lifted via a small magnetic field. The time dependence of the full system consists of a sudden switch in the hopping amplitude $v$ to a non-zero value.\\
The reference system, which is used as a starting point for the perturbation expansion, is modelled in the same way, but the hopping is switched to a different value. At this point we have to mention that although the auxiliary Hamiltonian used for preparing the initial state is correlated, an extension of the contour to imaginary times is not necessary. Here $G_{0}^{-1}$ has exactly the same form as Eq. \eqref{chap6:dis_action} and can therefore be treated on the contour depicted in Fig. \ref{chap6:fig:keldcont}. The reason for this is that the correlated impurity is decoupled from the bath for times smaller than zero. Consequently the density matrix $\rho_0$ splits into a direct product of a bath and impurity part. Fig. \ref{chap6:fig:short_time_plot} shows the time dependence of the occupation number on the interacting site. The black dotted curve is the exact time evolution of the full system, the blue curve is the time evolution of the reference system. The green and red data points show different expansion orders in the dual potential. The green points correspond to a dual theory without any diagram, the red curve to the solution including the first diagram.
Since this is a finite system, the solution is periodic (the period is larger than the time interval shown in the plot). The period of the oscillations of the magnetization is determined by the hopping amplitude $v$.
As one can see, both approximations improve the solution of the reference system towards the solution of the full system. In particular, already the zero-order solution corrects the oscillation period of the reference system.
The quality of the solution increases with the the order of approximation. The solution including the first diagram gives the best improvement. The superperturbation Keldysh theory is in quite good agreement with the exact result.\\
\begin{figure}[t]
\centering
\subfigure[]{
\includegraphics[scale=0.57]{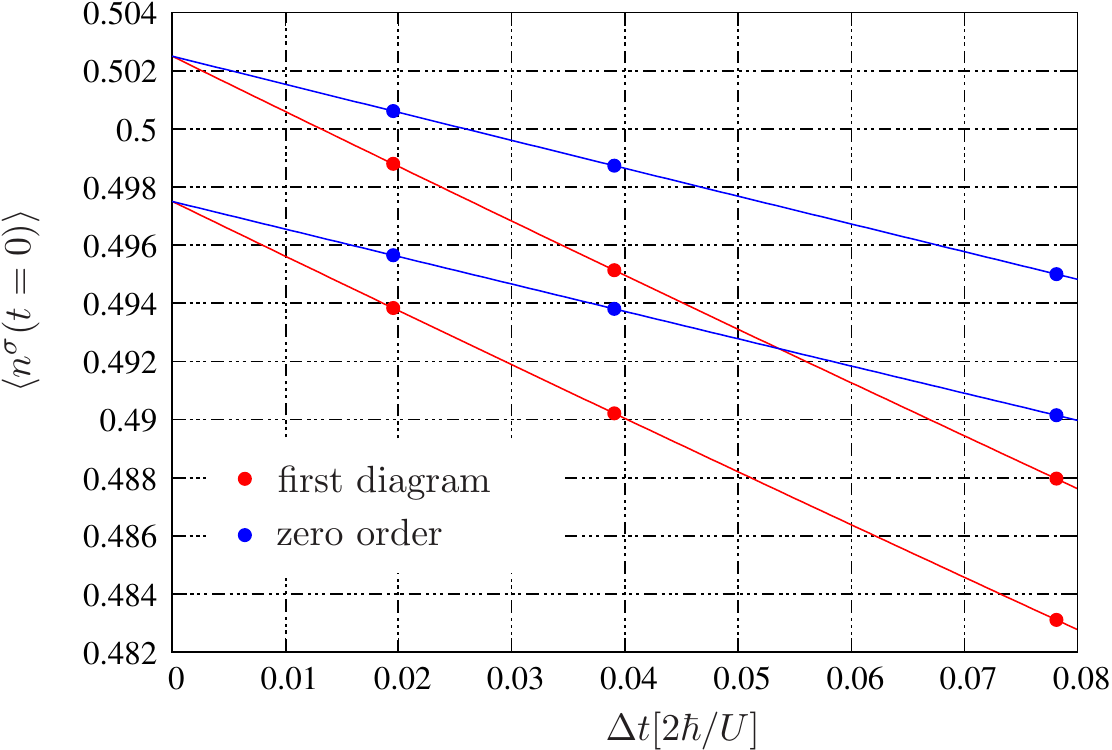}
\label{chap:6:fig:reg_t_0}
}
\subfigure[]{
\includegraphics[scale=0.57]{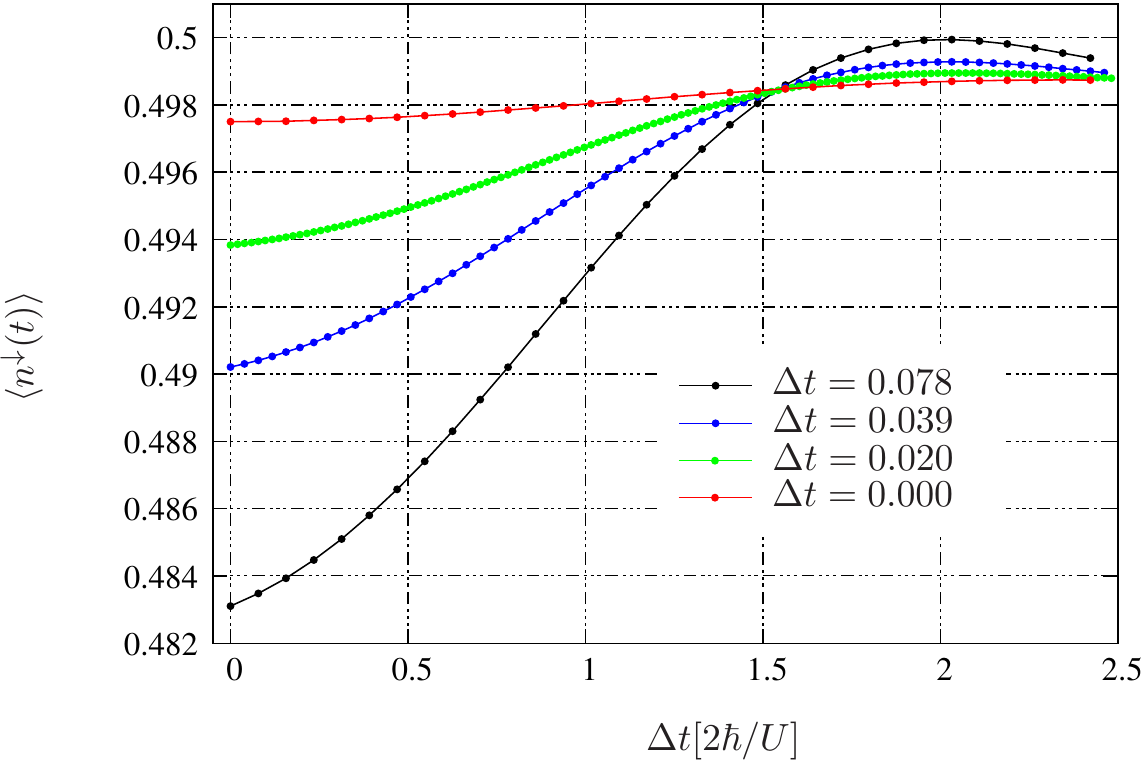}
\label{chap:6:fig:n_of_d_t}
}
\caption[Dependence of the final result on the grid size]{(Color online) Dependence of the final on the grid size. (a) Regression curve for the $n^{\sigma}(t=0)$ point of Fig. \ref{chap6:fig:short_time_plot}. A quadratic regression has been performed. (b) Plot of $n^{\sigma}(t)$ for different $\Delta t$. The dependence of the result on $\Delta t$ is quite strong. The reason for this behavior is that the effect of the initial magnetic field is very small, so that a high precision in the final result is needed to see this effect.}
\label{chap:6:fig:reg}
\end{figure}
In order to eliminate the systematic discretization error in the time argument, simulations for different grid sizes have been performed and the limit to a continuous time variable has been done numerically by quadratic regression. The final result can be expanded into a Taylor series in $\Delta t$ around the continuous solution:
\begin{equation}
n^{\sigma}(t,\Delta t)|_{\Delta t=0}\approx n^{\sigma}(t,0) + a\cdot \Delta t + b\cdot (\Delta t)^2\pm \dots .
\end{equation}
The three constants $a$, $b$ and the solution for a continuous time, $n^{\sigma}(t,0)$, have been calculated by quadratic regression. The procedure is depicted in Fig. \ref{chap:6:fig:reg_t_0} for the time point $t=0$. Here the dependence on $\Delta t$ is almost linear, but the quadratic regression is necessary to properly resolve the effect of the small initial magnetic field. The differences in $n^{\downarrow}(t)$ for different $\Delta t$ are shown in Fig. \ref{chap:6:fig:n_of_d_t}.
\subsection{Spin in fermionic bath}
An important non-trivial test for the method is the dissipation of a single spin, into and an infinitely long fermionic chain.
Although the solution of the reference system is necessarily periodic, for the superperturbation solution of the infinite system, this may not be so.

\begin{figure}[t]
\centering
\subfigure[]{
\includegraphics[scale=0.8]{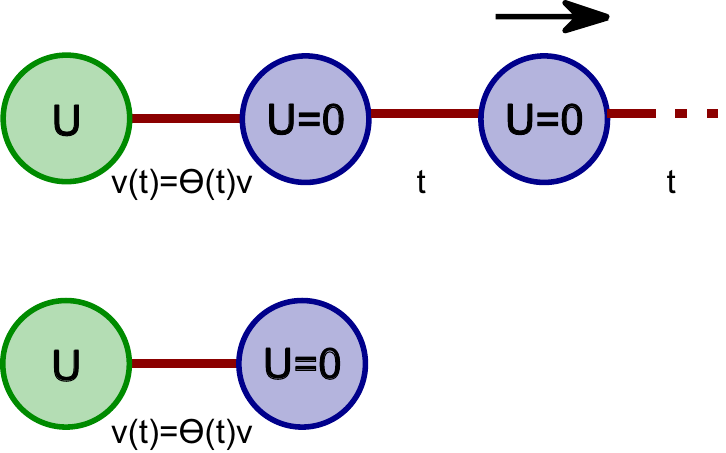}
\label{chap6:fig:model_spin_decay}
}\hspace*{0.8cm}
\subfigure[]{
\includegraphics[scale=0.7]{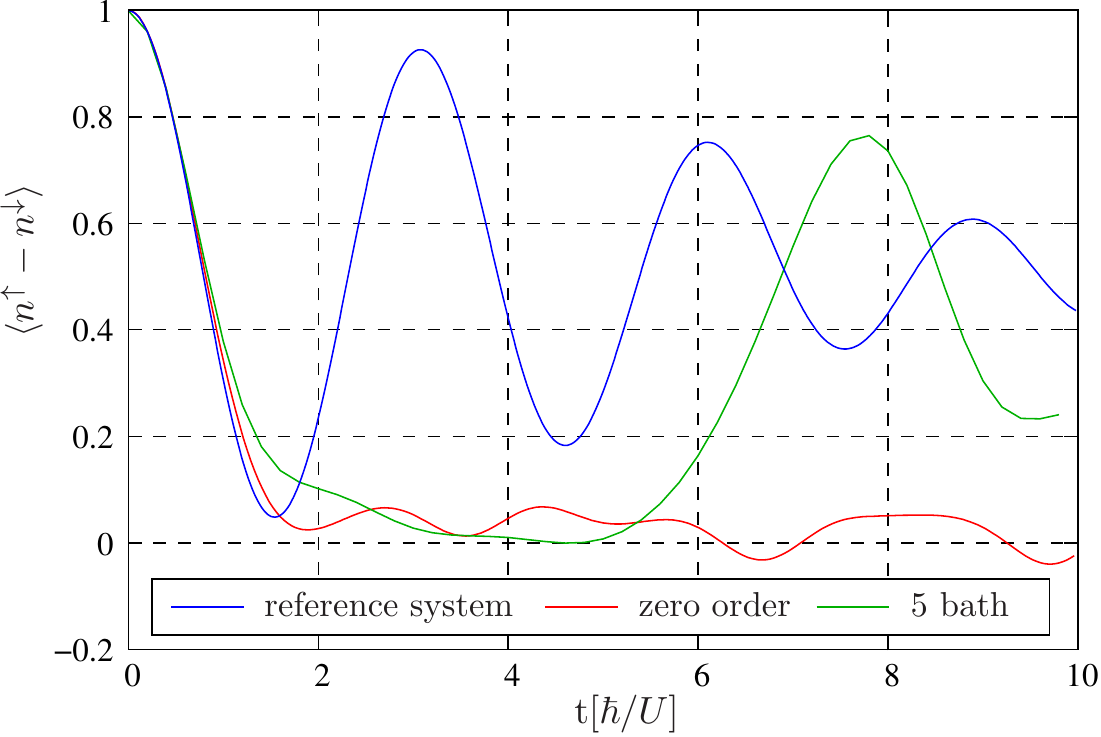}
\label{chap6:fig:spin_decay_nt}
}
\caption[]{(Color online) (a) An infinitely long chain with one interacting impurity at the left end is approximated by an effective two sites model with the same parameters. The initial state is prepared by a single spin on the impurity. The time-dependence is given by a sudden switch in the hopping to the rest of the chain at time zero. The parameters are $U=1$, $t=v=1$, $\beta=5$. (b) time-dependence of the magnetization for the two-site reference system (blue curve), in zero order dual perturbation theory (red curve) and of an impurity with the same parameters and a 5 site chain (green) curve.}
\label{chap6:spin_decay}
\end{figure}

The system under consideration consists of an isolated single magnetic Hubbard site  which is coupled to the infinite chain at time $t=0$ by a sudden switch in the hopping to the chain. Since the chain is infinite, the problem is a very simple example for a model coupled to an infinite fermionic reservoir. In the following we want to demonstrate that although the dual perturbation expansion starts from a Hamiltonian system, which obeys energy conservation, dissipation can be found already in the lowest order approximation, without  any dual diagram. This zero order approximation is very similar to a time-dependent Hubbard-I expansion (in the case $\tilde{\Delta}=0$, for the equilibrium case see e.g. Ref.\cite{Krivenko}) or a time-dependent variational cluster approach (t-VCA) (when $\tilde{\Delta}\neq0$, for equilibrium case see e.g. Ref.\cite{Potthoff})  and has the following very simple formulation:
\begin{equation}
G_{tt'}=\Bigl[ g^{-1}+(\tilde{\Delta}-\Delta)\Bigr]_{tt'}^{-1}.
\end{equation}
Here $G$ is the approximation for the Green's function of the full system and $g$ the Green's function of the reference system. For a chain the hybridization function $\Delta_{tt'}$ can be defined in terms of a continued fraction:
\begin{equation}
\Delta=V\Biggl(g_{0}^{-1}-V\biggl(g_0^{-1}-V\Bigl(g_0^{-1}-V\bigl(\dots\bigr)^{-1}V^{\dagger}\Bigr)^{-1}V^{\dagger}\biggr)^{-1}V^{\dagger}\Biggr)^{-1}V^{\dagger},
\end{equation}
where $g_{0}^{-1}$ is the non-interacting Green's function of a single fermionic site and $V$ a time-dependent hopping matrix. Both quantities have exactly the same form as shown in Example \ref{chap6:example:two_site}. The details of the model and results are shown in Figure \ref{chap6:spin_decay}. The reference system again consists of the same Hubbard impurity coupled to a single bath level.
At time zero a single spin is prepared on the impurity, then a hopping to the rest of the chain is switched on.
Figure \ref{chap6:fig:spin_decay_nt} shows the time-development of the magnetization on the impurity for the two site reference system (blue curve), in zero order dual perturbation theory (red curve) and of an impurity with the same parameters and a 5 site chain (green curve). At small times the length of the chain does not matter for the electron propagation, so all three curves coincide. At later times clear deviations are visible: The time-development of the reference system is governed by fast oscillations, which are caused by the reflection of the electron at the ends of the chain. The time-development of the magnetization for the model with 5 sites in the chain also shows a revival peak caused by the reflection of the electron, but this peak is found at later times, because of the longer chain.
The result of the perturbation expansion shows no backscattering peak and oscillates around zero, which is a clear indication that the presented approach allows to treat open systems starting from a perturbation around a closed Hamiltonian system.\\
It is clear that the perturbation expansion gives the best results, if the reference system is chosen in an optimal way, which essentially means that one should minimize a predefined matrix norm of $\Vert\tilde{\Delta}_{tt'}-\Delta_{tt'}\Vert$ by varying the effective parameters of the reference system. In the example at hand the minimum the effective hopping parameter of the reference system was chosen the same as the hopping parameter of the full system. The possibility to vary the reference system in principle allows to use the present solver as a solver for time-dependent DMFT calculations.

 \subsection{Comparison to CT-QMC}
 In the next step, we compare the dual approach to the recently developed CT-QMC method for non-equilibrium systems \cite{PhysRevB.79.035320}.
The underlying model is a single-level quantum dot with Coulomb interaction $U$.
The spin-degenerate level with orbital energy $\epsilon_{d}$ is coupled to two fermionic reservoirs ($L$, $R$) by a hybridization $V$.
The model Hamiltonian is given by
\begin{equation}
H=H^L+H^{LD}+H^{D}+H^{DR}+H^{R},
\end{equation}
where $H_D$ represents the isolated dot with dot operators $d^{(\dagger)}$:
\begin{equation}
H^{D}=\epsilon_d\sum_{\sigma=\uparrow, \downarrow} d^{\dagger}_{\sigma} d_{\sigma} + U d^{\dagger}_{\uparrow} d_{\uparrow}  d^{\dagger}_{\downarrow} d_{\downarrow}.
\end{equation}
Furthermore, $H^{L}$ and $H^{R}$ describe a free electron gas in the left and right reservoir
\begin{equation}
H^{L/R}=\sum_{\vec{k}\sigma} (\epsilon_{\vec{k}\sigma} - \mu_{L/R}) c^{\dagger}_{\vec{k}\sigma}c_{\vec{k}\sigma},
\end{equation}
where $c^{(\dagger)}_{\vec{k}\sigma}$ are creation and annihilation operators for the conductance electrons with momentum $\vec{k}$ and spin $\sigma$. The energy distribution of the electrons in the two reservoirs obeys the Fermi function $f(E)=[1+\exp(\beta (\epsilon_{k}-\mu_{L/R})]^{-1}$, assuming each reservoirs to be in thermal equilibrium with corresponding electro-chemical potentials $\mu_L$ and $\mu_R$. A bias voltage $V$ can be applied to the dot by shifting the potentials $\mu_{L/R}$. However, in the following we consider only the case $V=0$ with $\mu_{L,R}=0$.
The dot is symmetrically coupled to the left and right leads by the spin-conserving tunneling Hamiltonian
\begin{equation}
H^{LD/DR}=\sum_{\vec{k}\sigma }V_{\vec{k}} c^{\dagger}_{\vec{k}\sigma}d_{\sigma}+V^{\ast}_{\vec{k}} d^{\dagger}_{\sigma}c_{\vec{k}\sigma}.
\end{equation}
The strength of the tunneling can be characterized by the constant $\Gamma = 2 \pi \rho \vert V \vert^2$, where $\rho$ is the density of states in the leads. The density of states is taken to be constant with a hard cutoff \cite{PhysRevB.79.035320} at $\epsilon = \pm 10 \Gamma$. In what follows, the energy scales are defined by the level broadening $\Gamma$.

We introduce the reference system for the dual approach resembling the original model but reducing the leads to a single electronic level
\begin{equation}
\tilde{H}^{L/R}= \tilde{\epsilon}\sum_{\sigma } c^{\dagger}_{\sigma}c_{\sigma}.
\end{equation}
The energy of the left and right lead $\tilde{\epsilon}$ is chosen to be equal to the orbital energy of the dot.
For the model we investigated, the best agreement with MC is achieved when both chemical potentials $\tilde{\mu}_{L/R}$ in the auxiliary system equal the value of the level energy $\tilde{\epsilon}$. In contrast, the chosen level energy of the axillary system related to the level energy of the dot is of minor importance.
\begin{figure}[t]
\centering
\subfigure[]{
\includegraphics[scale=0.8]{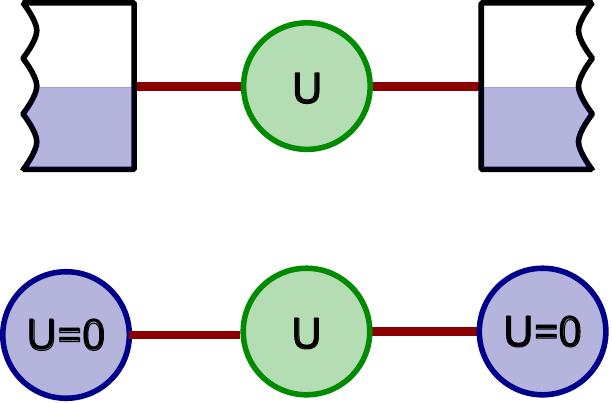}
\label{chap6:fig:model_comp_ctqmc}
}\hspace*{0.8cm}
\subfigure[]{
\includegraphics[scale=0.8]{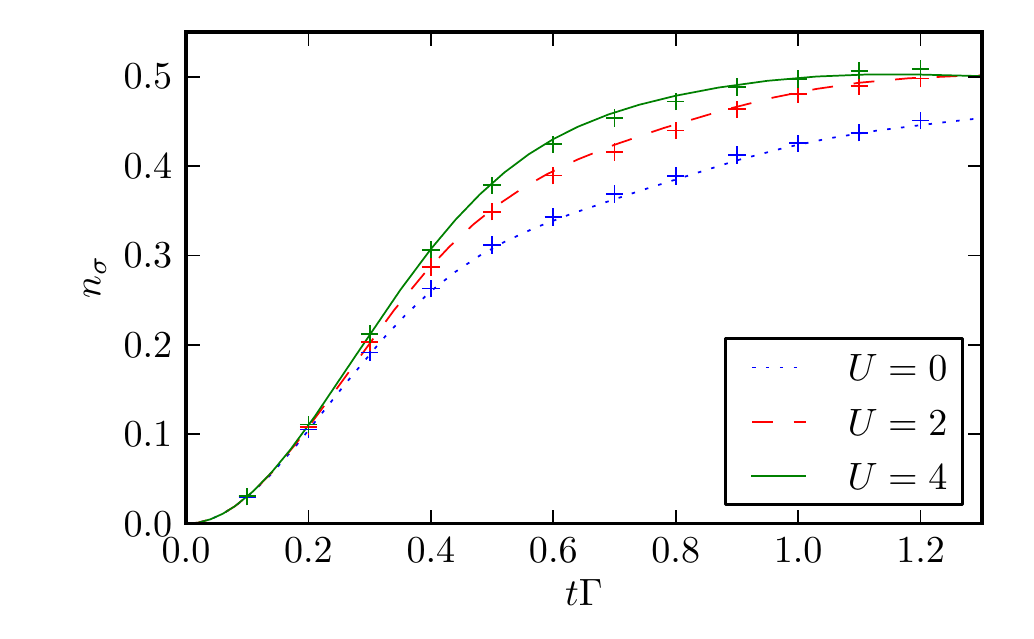}
\label{chap6:fig:comp_ctqmc}
}
\caption[]{(a.) Model under consideration: a quantum dot (interacting impurity) is coupled to two infinite reservoirs at half filling. The time-dependence is given by a sudden switch in the tunneling to the reservoirs to a non-zero value at $t=0$. The reference system is a single impurity with the same interaction, but a bath which consists of two additional sites only. (b.) Comparison of the occupation of the dot in lowest order dual approximation (lines) to CT-QMC data (symbols) for various values of $U$. The temperature is $\beta\Gamma=10$.}
\label{chap6:comp_ctqmc}
\end{figure}
Fig. \ref{chap6:comp_ctqmc} shows the results of the dual approach in lowest order (lines) for different values of $U$ compared to data obtained by the diagrammatic Monte Carlo method (symbols). As the non-equilibrium Monte Carlo method suffers from the dynamical sign problem we restrict ourselves to the short-time dynamics of the system. The electron density was calculated for the specific parameters leading to the half filled case with $\epsilon_d = - \frac{U}{2}$ and temperature $\beta \Gamma = 10$. Initially, the dot is empty and the coupling between the dot and the leads is switched on at $t=0$. The comparison between the dual approach and non-equilibrium Monte Carlo shows good agreement for the chosen parameter range.
For very strong interaction and large times, however, higher order corrections of the dual perturbation series become significant.

\section{Conclusions}
In the presented work we have generalized the dual fermion method to treat strongly correlated systems out of equilibrium.
We have formulated the dual perturbation theory on the Keldysh contour and described the implementation of lowest order dual-diagram for a quite broad class of physical non-equilibrium problems. We also presented the details of the numerical implementation within the exact diagonalization scheme. First tests show that the method works as well for finite as for open systems in a fermionic bath. The long timescale non-equilibrium propagation proved to be accessible within the dual fermion perturbation theory, which is an advantage compared to non-equilibrium CT-QMC based methods. We also considered a simpler version of the presented method (a zero-order approximation), the time-dependent variational cluster approximation, that demands much less numerical effort and seems to be a promising tool for more complex systems. The dual fermion non-equilibrium scheme allows to go beyond zero-order and include physically important diagrammatic series which can describe the long time scale propagation of quantum systems out of equilibrium.
\section{Acknowledgments}
We thank M. Eckstein, I. Krivenko, M. Balzer, M. Potthoff, Ph. Werner, and A. Cavalleri for valuable discussions. This work has been supported by the Cluster of Excellence "Nanospintronics" (LExI Hamburg) and DFG Grant(436113/938/0-R).
\bibliographystyle{apsrev4-1}
\bibliography{water_s}
\end{document}